\documentclass[aps,prd,twocolumn,groupedaddress,showpacs,showkeys,nofootinbib,amssymb]{revtex4}
\usepackage[latin1]{inputenc}
\usepackage{graphicx}
\usepackage[english]{babel}
\usepackage{amsmath}
\usepackage{amssymb}
\usepackage{amsfonts}
\usepackage{colordvi}
\usepackage{psfrag}
\usepackage{color}
\usepackage{tipa}
\begin{document}
\def\pp{{\, \mid \hskip -1.5mm =}}
\def\cL{{\cal L}}
\def\be{\begin{equation}}
\def\ee{\end{equation}}
\def\bea{\begin{eqnarray}}
\def\eea{\end{eqnarray}}
\def\beq{\begin{eqnarray}}
\def\eeq{\end{eqnarray}}
\def\tr{{\rm tr}\, }
\def\nn{\nonumber \\}
\def\e{{\rm e}}

\title{\textbf{Astrophysical structures from primordial quantum black holes}}

\author{Salvatore Capozziello$^{1,2}$, Gerardo Cristofano$^{1,2}$, Mariafelicia  De Laurentis$^{1,2}$}

\affiliation{\it $^{1}$Dipartimento di Scienze Fisiche, Università
di Napoli {}``Federico II'', $^{2}$INFN Sez. di Napoli, Compl. Univ. di
Monte S. Angelo, Edificio G, Via Cinthia, I-80126, Napoli, Italy}

\date{\today}

\begin{abstract}
The characteristic sizes of astrophysical  structures, up to the whole observed Universe, can be recovered, in principle,  assuming that   gravity is the overall
interaction assembling  systems starting from   microscopic scales, whose order
of magnitude is  ruled by the Planck length and the related Compton wavelength.  This result agrees with the absence of screening
mechanisms for the gravitational interaction and could be connected to the presence of Yukawa corrections in the
Newtonian potential which introduce typical interaction lengths. This result directly comes out from  quantization of primordial black holes and then characteristic interaction lengths directly emerge from quantum field theory.

 \end{abstract}

\maketitle

\section{Introduction}
\label{uno}

Framing the astrophysical, self-gravitating structures in a self-consistent scheme is one of
the main puzzles of modern physics since the growing amount
of observations seems to escape any coherent interpretative  scheme.
Essentially, from the fundamental physics point of view, we would
like to re-conduct cosmic structures and their evolution to some
unifying theory in which all the today observed interactions are
treated under the same standard. In such a scheme, what we observe
today on cosmological and astrophysical scales would be just a
consequence of quantum fluctuations in early epochs. Then, we
should seek for some dynamical  mechanism that, after one (or
more than one) symmetry breaking, would be capable of yielding
structures like clusters of galaxies, galaxies and then stars from
primordial quantum spectra of perturbations.

The so called ``inflationary paradigm'' \cite{kolb} related to
several unifying theories (e.g. superstrings, GUT, SUSY) seems adequate to this goal if one were able to
definitely select the right model.

On the other hand, particle physics needs cosmological
predictions and observations since the energies for testing
unified theories are so high that it is extremely unlikely that they
will be ever reached on Earth--based laboratories.
As a matter of fact, cosmology needs particle physics and vice
versa. The point is that remnants of primordial epochs should be
found by cosmological observations and, by them, one should
constrain elementary particle physics models \footnote{Results of Large Hadron Collider (LHC) at CERN could be the "experimentum crucis" since, due to the energies involved, their outcome should be related to the thermodynamical state of primordial Universe.}.

This philosophy has been pursued by several researchers; first of
all by Sakharov \cite{sakharov} who in 1965 argued that quantum
primordial fluctuations should have expanded towards the present
epoch leading first to classical energy--density perturbations
and, after the decoupling from the cosmological background, to the
observed galaxies, clusters, super-clusters of galaxies, and
afterwards stars. Shortly, the underlying issue of any modern
theory of cosmological perturbation is this: primordial quantum
fluctuations should be enlarged by cosmological dynamics to the
present large scale structures. Now the problem is not only
whether observations agree with this scheme (e.g. COBE, IRAS, WMAP, PLANCK
data or large scale structure surveys \cite{hucra}) but, mainly,
whether the astrophysical and cosmological systems ``remember''
their quantum origin or not.

Despite of the apparent sharp division of the classical and
quantum worlds, macroscopic quantum phenomena exist and some
behaviors of classical systems can be explained only in the
framework of quantum mechanics. The high $T_{c}$
superconductivity and several other macroscopic coherent systems
 are interesting instances of these peculiar
phenomena in which a quantum ``memory'' persists at the
macroscopic scale.

Some intriguing conjectures have been proposed to find quantum
signatures  at the classical, macroscopic scales of astrophysical structures
\cite{calogero}.
In this scheme, classical laws of force  describing the
interactions among the constituents of $N$--particle systems of
mean length scale $R$, lead to a quantum  characteristic action
per particle. The forces considered can be, for instance, the
electromagnetic interactions between charged particles in large
macroscopic systems as charged beams in particle accelerators,
plasmas, and neutral dipolar crystals,
 or the strong interactions between quarks in hadronic bound aggregates,
 and so on.

Having such a procedure at hand, it seems very natural  to
investigate whether it can be applied to determine the existence
of quantum  signatures or ``memories'' for astrophysical
structures.

It can be shown that the
order of magnitude of characteristic observed radii of typical
galaxies and large scale structures can be inferred starting from the microscopic
fundamental scales through a phenomenological scaling law \cite{scott1,scott2}.
Further scaling relations can be derived, showing that such a
connection can be equivalently  obtained either by considering the
nucleons as the elementary constituents of a  galaxy, or, as usual
in astrophysics, by considering stars as its granular components.
The key issue is that we can define an interaction length
$\lambda$ and a typical number of elementary constituents $N$
(e.g. $N_n$ the number of nucleons with $\lambda_c$, the Compton
wavelength,  or $N_s$ the number of stars with $\lambda_s$, the
typical interaction length of a star, for example the ''size" of
the Solar System) and, by the relation

\beq\label{compton} R\simeq\lambda\sqrt{N}\,,\eeq to obtain the
observed typical size (in the case of galaxies we obtain $\sim
10$ kpc a typical galactic scale).

In the present paper, we show that such a relation holds for all
self-gravitating systems as stellar globular clusters, clusters
and super-clusters of galaxies and we show that it can be
implemented by taking into account a gravitational interaction,
explicitly scale-dependent, as predicted by
quantum field theory in the low energy limit and resulting, at macroscopic level, as an extended theory of gravity where more general actions of curvature invariants than that by Hilbert and Einstein are considered.

More precisely, renormalizable quantum fields  in a  gravitational background
give rise to  modifications, in the low energy limit, of  the Newtonian
potential. Furthermore, if we do not require enormous amounts of
dark matter as the only mechanism to explain the puzzle of the
present day astrophysical observations, a scale--dependent
gravitational interaction is  needed. This point of view is
supported by several authors \cite{mcgaugh} which state that
recent experiments on cosmic microwave background as BOOMERANG, WMAP and PLANCK
\cite{debernardis} or astrophysical structures can be explained
by taking into account modified Newtonian dynamics
\cite{mnras}.

The present paper is organized as follows. Sec. II is devoted to the discussion of stationary points emerging from quantum field theory in curved space-time where Schwarzschild radius and fundamental electric charge are considered. In Sec. III, we discuss the Lagrangian approach to the so called "attractor mechanism"
while, in  Sec. IV, we deal with the
emergence of quantum charged black holes which could give  rise, in principle,  to the fundamental scale ruling all the further, larger astrophysical  scales.
 In Sec. V, we show that the  weak field limit of extended theories of gravity, coming from quantum field theories formulated  in curved space-times,
gives rise to Yukawa-like corrections to the Newtonian
potential, which, naturally imply characteristic lengths. Sec. VI
is devoted to the discussion of  characteristic sizes of
astrophysical structures. In Sec. VII, we discuss how
time-statistical fluctuations of granular components of
self-gravitating systems are related to their characteristic
sizes. Conclusions are drawn in Sec. VIII.

\section{Characteristic lengths from quantum field theory}
\label{due}
 Let us  start  our considerations from a  Reissner-Nordstrom metric  in order
to identify  a space-time with a unitary electric charge and a mass acting as source of gravitational field. It is
\cite{chandrasekhar}
\begin{equation}
c^2dt^2=\left(1-\frac{r_S}{r}+\frac{r_Q}{r^2}\right)c^2dt^2-\frac{dr^2}{1-\frac{r_S}{r}+\frac{r_{Q}^{2}}{r^2}}-r^2d\Omega^2\, ,
\label{metrica}
\end{equation}
where $d\Omega=d\theta^2+\sin^2\theta d \phi^2$ is the solid angle, $r_S$  is Schwarzschild radius of the massive body $\displaystyle{r_S=\frac{2GM}{c^2}}$
where $G$ is the gravitational constant, and $r_Q$ is a length-scale related to the electric charge $Q$  $\displaystyle{r_{Q}^2=\frac{Q^2G}{4\pi\epsilon_0 c^4}}$,
where $\displaystyle{\frac{1}{4\pi\epsilon_0}}$ is the Coulomb  constant. In the limit where the charge $Q$ (or equivalently, the length-scale $r_Q$) goes to zero, one recovers the Schwarzschild metric. The classical Newtonian theory of gravity is recovered in the limit  $\displaystyle{\frac{r_S}{r} \rightarrow 0}$. In this limit, the  Minkowski metric
\begin{equation}
c^2d\tau^2=c^2dt^2-dr^2-r^2d\Omega^2
\label{mink}
\end{equation}
is recovered.
Although charged black holes  with $r_Q<<r_S$  are similar to the Schwarzschild black hole, they have two horizons: the event horizon and an internal Cauchy horizon. As usual, the event horizon can   be reliably identified by analyzing the equation
\begin{equation}
g_{00}=1-\frac{r_S}{r}+\frac{r_Q}{r^2}=0\, .
\label{gtt}
\end{equation}
This quadratic equation for $r$ has the solutions
\begin{equation}
r_{\pm}=\left(\frac{\sqrt{G}}{c^2}\right)\left(\frac{r_S\pm \sqrt{r_{S}^2-4r^2_{Q}}}{r}\right)\, .
\label{r}
\end{equation}
These concentric event horizons become degenerate for $2r_Q = r_S$ which corresponds to an extremal black hole \footnote{Black holes with $2r_Q > r_S$ are believed not to exist in nature because they would contain a naked singularity; their appearance would contradict cosmic censorship hypothesis which is generally believed to hold \cite{Shapiro}}. That is
\begin{equation}
\frac{2GM}{c^2}=\frac{2Q\sqrt{G}}{c^2\sqrt{4\pi\epsilon_0}}
\label{exbh}
\end{equation}
or better squaring both members get
\begin{eqnarray}
\frac{4G^2M^2}{c^4}&=&\frac{2Q}{4\pi\epsilon_0}\frac{G}{c^4}\, ,\nonumber\\
GM^2&=&\frac{Q^2}{4\pi\epsilon_0}\, .\nonumber\\
\label{exbh1}
\end{eqnarray}
If the extreme black hole has also a magnetic charge $m$ then
\begin{equation}
r_{Q}^2\rightarrow r_{Q}^2+r_{m}^2=\frac{Q^2}{4\pi\epsilon_0}\frac{G}{c^4}+m^2\frac{\mu_0}{4\pi}\frac{G}{c^4}
\label{exbh2}
\end{equation}
and going through the same steps, we get
\begin{equation}
GM^2=\frac{Q^2}{4\pi\epsilon_0}\frac{G}{c^4}+m^2\frac{\mu_0}{4\pi}\frac{G}{c^4}\,,
\label{exbh3}
\end{equation}
or, in Planck units\footnote{$G=c=\hbar=(4\pi\epsilon_0)^{-1}=1$},
\begin{equation}
M^2=Q_{e}^2+Q_{m}^2\,,
\label{exbh4}
\end{equation}
after using the more convenient notations $Q_e=Q$ and $Q_m=m$ for the electric and magnetic charges.
Some comments are in order at this point.
\begin{itemize}
\item {Eq.(\ref{exbh4})  represents a no-net  force condition: black holes appear to be like {\it dyons} that are objects with mass $M$, electric charge $Q_e$ and magnetic charge $Q_m$. Then the above relation tells that the attractive gravitational force between  such two objects at distance $r$ compensates the repulsive electric and magnetic forces between them}.
\item { Eq.(\ref{exbh4}) expresses the saturation of the so called Bogomol'nyi-Prasad-Sommerfeld (BPS)  bound (see \cite{Ferrara}) expressing the requirement that the local charge density does not exceed the local matter density  (this is also called  the "dominant energy condition").}
\end{itemize}
In conclusion, these considerations point out that it is always possible to select fundamental length scales by considering masses and charges. Our aim is now to derive fundamental scales from a field theory and then connect them to astrophysical self-gravitating systems.

\section{The Lagrangian approach}
\label{tre}

The above relation can be
derived from a Lagrangian formulation of extremal black holes.
Basically there are two  approaches,  a non-supersymmetric one (see for example \cite{Goldstein})
 and a supersymmetric one (see for example \cite{Ferrara}.
In both approaches there are  scalar fields (also called "moduli")
coupled to gauge fields with dilaton like coupling. The scalar quantities are considered, at the beginning,  without
self-interacting  potential.

For the non-supersymmetric case,
let us consider a general situation with a 4-dimensional Lagrangian with $U(1)$ gauge fields and scalars with the "inclusion" of "axion" type couplings (see the last term) of the form
\begin{eqnarray}
{\cal S}&=&\frac{1}{k^2}\int dx^{4} \sqrt{-g}\left[ R-2(\partial \varphi_{i})^2+ \right.\nonumber\\ &&
-f_{ab}(\varphi_{i})F^{a}_{\mu\nu}F^{b\mu\nu}+\nonumber\\ &&
-\left.\frac{1}{2}{\tilde f}_{ab}(\varphi_i)F^{a}_{\mu\nu}F^{b}_{\rho\sigma} \epsilon^{\mu\nu\rho\sigma}\right]\nonumber\\
\end{eqnarray}
which has $i=1,...,N$ scalars with standard kinetic energy terms and $a=1,...,M$ gauge fields \footnote{ A more general kinetic energy term for the scalars would arise from a
Kahler  supersymmetric theory. However,  it is not necessary for our considerations
here. }.
 It can be shown that the fields appearing in the metric obey equations of motion governed by  an effective potential

\begin{eqnarray}
V_{eff}(\varphi_i)&=&f^{ab}\left(Q_{e\,a}-{\tilde{f}}_{ac}Q_{m}^{c}\right)\left(Q_{e\, b}-{\tilde{f}}_{bd}Q_{m}^{d}\right)\nonumber\\ &&
+{\tilde{f}}_{ab}Q_{m}^{a}Q_{m}^{b}\,.
\label{Veff}
\end{eqnarray}
which is proportional to the energy density in the electromagnetic field. $Q_e$ and $Q_m$  are the electric and magnetic charges associated to the gauge field $F^a$,
$f^{ab}$ being the inverse of the gauge coupling function $f_{ab}$.
An effective potential can be derived also for the  supersymmetric case. Before defining this, we observe that in both approaches, the so called {\it "attractor mechanism"}  takes place once the following conditions are satisfied:

\begin{itemize}

\item
{the derivative of the potential has to be}
\begin{eqnarray}
\partial_{i}V_{eff}(\phi_{i0})=0
\end{eqnarray}
{where$\phi_{i0}$ are the critical field values;}

\item
{the positive eigenvalue condition}
\begin{eqnarray}
M_{ij}=\frac{1}{2}\partial_{i} \partial_{j} V_{eff}(\phi_{i0})>0
\end{eqnarray}
 {has to hold  (see \cite{Trivedi} for the simplest system described by one scalar and two gauge fields).}
\end{itemize}
In the  $N=2,4$ supersymmetric theories the above conditions are automatically satisfied.
In fact,  let us consider the effective potential  for a black hole $V_{BH}$ with electric and magnetic charges in a supergravity $D=4,N=4$ theory (see \cite{Ferrara})

\begin{eqnarray}
V_{BH}(\phi,a,Q_m,Q_e)&=&e^{2\phi}\left[\left(a+ie^{-2\phi}\right)Q_m-Q_e\right]\times\nonumber\\&&\left[\left(a-ie^{-2\phi}\right)Q_m-Q_e\right]\label{Ve}
\end{eqnarray}
where $\phi$ is the dilaton field, $a$ the axion one and only one component of the electromagnetic charges is considered for simplicity. Combining terms, one gets

\begin{eqnarray}
V_{BH}=e^{2\phi}(Q_e-aQ_m)^2+e^{-2\phi}Q^2_{m}\nonumber\\
\label{VBH}
\end{eqnarray}
On the other hand from conformal field theory (CFT) approach \footnote{The part of the interaction potential between charges dependent on the positions can be omitted due to the exat cancellation with its gravitational counterpart after imposing the no-net force condition (\ref{exbh4} or its generalization which we will derive in (\ref{29}).
} \begin{eqnarray}
V_{eff}^{CFT}=R_{c}^{2}\left(Q_e-\frac{\theta}{2\pi}Q_m\right)^2+\frac{1}{R^2}Q^{2}_{m}
\end{eqnarray}
where $R_c$ is the compactification radius of the scalar Fubini field \cite{fubini}.
If we equate $V_{eff}^{CFT}$ with $V_{BH}$ get the following identifications

\begin{equation}R_{c}^{2}=e^{2\phi}\, ,\qquad \frac{\theta}{2\pi}=a\,,\end{equation}
where $\theta$ is a  parameter (see \cite{Gthooft})
and the dilaton axial scalar is
\begin{equation}
s=a+ie^{-2\phi}=\frac{\theta}{2\pi}+\frac{i}{R_{c}^2}\,.
\end{equation}
Using the variable
\begin{equation}
\zeta=\frac{\theta}{2\pi}+i\frac{1}{g}
\end{equation}
we can see that
\begin{eqnarray}
V_{eff}=g\left(Q_e-\frac{\theta}{2\pi}Q_m\right)^2+\frac{1}{g}Q_{m}^2
\end{eqnarray}
is invariant under the generalized duality transformations $\zeta\rightarrow\zeta+1$ and $\zeta=-\frac{1}{\zeta}$ \cite{cardi}.

Let us derive now the stationary (or critical ) points of the above dynamics.
We start with
\begin{eqnarray}
V_{eff}=R_{c}^2\left(Q_e-\frac{\theta}{2\pi}Q_m\right)^2+\frac{Q_{m}^2}{R_{c}^2}\label{I}\,.
\end{eqnarray}
Imposing $\displaystyle{\frac{\partial V_{eff}}{\partial R_c}=0}$, we  get \footnote{The derivatives of $V_{eff}$ are taken with respect to $R_c$ and not $\phi$ only for the sake of  simplicity, since results (\ref{II}) and (\ref{III}) are not altered.}
\begin{eqnarray}
2R_c\left(Q_e-\frac{\theta}{2\pi}Q_m\right)^2-2\frac{Q_{m}^2}{R{c}^3}=0
\end{eqnarray}
\begin{eqnarray}
\Rightarrow R_c\left(Q_e-\frac{\theta}{2\pi}Q_m\right)^2=\frac{Q_{m}^2}{R_{c}^3}
\end{eqnarray}
\begin{eqnarray}
\Rightarrow R_{c}^4=\frac{Q_{m}^{2}}{Q_e-\frac{\theta}{2\pi}Q_m}
\end{eqnarray}
\begin{eqnarray}
\Rightarrow R_{H}^2=\frac{Q_{m}}{Q_e-\frac{\theta}{2\pi}Q_m}=e^{2\phi_{H}}\label{II}
\end{eqnarray}
where $R_H$ and $\phi_H$ indicate the corresponding values at the
horizon of the black hole. Furthermore, we have
\begin{eqnarray}
\frac{\partial^{2} V_{eff}}{\partial R^{2}_{c}}&=& 2\left(\frac{Q_{m}}{Q_{e}-\frac{\theta}{2\pi}Q_m}\right)^2+6\frac{Q_{m}^{2}}{R_{c}^{4}}\mid_{R_{c}=\frac{Q_{m}}{Q_e-\frac{\theta}{2\pi}Q_m}}\nonumber\\&&
=8\left(\frac{Q_{m}}{Q_e-\frac{\theta}{2\pi}Q_m}\right)^2>0
\end{eqnarray}
This means that  the two conditions for the attractor mechanism are satisfied and the  mass $M$ and charges $Q_e,Q_m$ of the extremal black holes saturate the BPS bound,
 \begin{eqnarray}
 M^2=e^{2\phi_{H}}\left(Q_e-\frac{\theta}{2\pi}Q_m\right)^2+e^{-2\phi_{H}}Q_{m}^{2}\,,\label{III}
 \end{eqnarray}
 or
  \begin{eqnarray}
  M^2=R^{2}_{H} \left(Q_e-\frac{\theta}{2\pi}Q_m\right)^2+\frac{Q_{m}^2}{R^{2}_{H}}\, ,\label{29}
 \end{eqnarray}
 that generalize the above Eq.(\ref{exbh4}).
 After substituting  $\displaystyle{R^{2}_{H}=\frac{Q_{m}^{2}}{Q_e-\frac{\theta}{2\pi}Q_m}}$, we get
  \begin{eqnarray}
  M^2=2 \left(Q_e-\frac{\theta}{2\pi}Q_m\right)Q_m\, ,
  \label{IV}
  \end{eqnarray}
 Let us now briefly comment on the above relations (\ref{I}),(\ref{II}),(\ref{III}),(\ref{IV})
 \begin{itemize}
 \item In the metric approach (see Sec.\ref{due}),  the relation (\ref{III}) is a consequence of $g_{00}=0$, that is
 we can say that the condition $g_{00}=0$  points out  a phase transition, that is charged black holes are forming with mass $M$ and charges $Q_e,Q_m$ obeying the relation (\ref{III}).
 {We can imagine that, at its very beginning, the Universe was constituted by these primordial and "quantum" black holes with  masses $M_i=M$, charges, $Q_{e}^{i}=Q_{e}$ and  $Q_{m}^{i}=Q_{m}$ whose stability condition reads as
  \begin{eqnarray}
\sum_{i,j} M_i M_j&=&\sum_{i,j} e^{2\phi}\left(Q_{e}^{i}-\frac{\theta}{2\pi}Q_{m}^{i}\right)\times\nonumber\\ &&\left(Q_{e}^{j}-\frac{\theta}{2\pi}Q_{m}^{j}\right)+\left[e^{-2\phi}Q_{m}^{i}Q_{m}^{j}\right]\,,\nonumber\\
   \end{eqnarray}
that is, using $M_i=M_j=M$, $Q_{e}^{i}=Q_{e}^{j}=Q_e$ and $Q_{m}^{i}=Q_{m}^{j}=Q_m$  as in Eq. (\ref{III}). We can interpret  the relation (\ref{II}) as the  compactification radius of the Fubini scalar field \cite{fubini} describing the "correlations" of the black holes.   Such scalar fields can be interpreted  as the primary fields of the conformal field theory (CFT) describing the phase transition points or  the "critical" points.
 Notice that the same relation  (\ref{II}) is realized in the CFT description of a quantum Hall fluid between the Fubini scalar field compactification radius and the electric and magnetic charges of the relavant excitations of the Hall fluid  \cite{hall} .}
 \end{itemize}
 We have now all the ingredients to fix fundamental characteristic lengths from quantum black holes and then derive astrophysical sizes.

 \section{Quantum Black Holes to fix the characteristic lengths}
 \label{quattro}
Without loosing any generality, let us consider  the axion field
$\displaystyle{a=\frac{\theta}{2\pi}=0}$ and write
  \begin{eqnarray}
  M^2=2Q_e Q_m
  \label{V}
   \end{eqnarray}
   In order to consider such black holes as quantum objects, we use the Dirac quantization condition \cite{diracquan}
  \begin{eqnarray}
  2Q_e Q_m=n \hbar c
   \label{VI}
   \end{eqnarray}
   with $n$ an integer $>0$\footnote{Notice that standard units have been introduced here.}.
   Substituting into the previous relation, one gets
    \begin{eqnarray}
    GM^2=n\hbar c
    \label{VII}
   \end{eqnarray}
   For $n=1$, we get the lowest mass allowed for a quantum black hole (that we consider a primordial black hole):
   \begin{eqnarray}
   M_{BH}=\sqrt{\frac{\hbar c}{G}}=M_{Planck}.
   \end{eqnarray}
 Let us comment on this  result. At  very beginning ($t\simeq10^{-43}$ sec) and at  the "Planck" temperature, charged black holes were formed and constituted all the Universe mass. Thanks to the Dirac quantization condition (\ref{VI}), relation (\ref{VII}) appears to be very general due to the precence of the factor $n \hbar$, describing the angular momentum of the mass $M$.    It is easy   to show that such a quantization relation is valid for any given astrophysical mass by considering the ratio
 \begin{eqnarray}
 n_{astro}=\frac{r_{S-astro}}{2\lambda_{Compton-astro}}\,.
   \end{eqnarray}
   where we have assumed the Schwarzschild radius and the  Compton length for any astrophysical structure
We can  check the scaling hypotesis in a few steps. We have

 \begin{eqnarray}
 {GM_{Planck}^{2}}=n_{Planck}\hbar c\,,
    \end{eqnarray}
 where $n_{Planck}=1$  and
     \begin{eqnarray}
 {GM_{u}^{2}}=n_{u}\hbar c\, ,
    \end{eqnarray}
    where $M_{u}$ is the whole Universe mass.
Taking their ratio, one gets
 \begin{eqnarray}
 \frac{n_u}{n_{Planck}}=\left(\frac{M_u}{M_{Planck}}\right)^2
    \end{eqnarray}
  By using $M_u = n_p m_p$ with $m_p=10^{-27}$Kg, the proton mass
 and $n_p=10^{78}$,  the number of  protons in the Universe, we get

   \begin{eqnarray}
\Rightarrow n_{u}=(10^{60})^2=10^{120}\,,
    \end{eqnarray}
 That is the quantum relation (\ref{VII}) applies, in principle,  to small scale structures (primordial Black Holes)  up to the whole Universe, with $120$ orders of magnitude between them, just the order of magnitude between the value of the cosmological constant $\Lambda$ at Planck scales and its present value ( \cite{weinberg}).

We can rewrite relation (\ref{VII})
 in another interesting form, using  $n\hbar =J$, the angular momentum of the mass $M$ as
   \begin{eqnarray}
   J=\frac{G}{c}M^2\label{J}\,.
   \end{eqnarray}
Notice the strong resemblance with the corresponding relation for string states \cite{green}.
In order to show how such a scaling relation applies to any self-gravitating system, relation (\ref{J}) can be rewritten in a more suitable form as
\begin{equation}
J=\left(\frac{G M^{2}}{R}\right) \left(\frac{R}{c}\right)=ET\label{J1}\, ,
\end{equation}
where $E$ is the characteristic gravitational energy of the astrophysical structure of "size" $R$ and $T$ is the characteristic crossing time for a light beam. Relation (\ref{J}) is the starting point for the discussion in Sec. \ref{sette}.

 \section{The emergence of characteristic interaction lengths}
 \label{cinque}

The above discussion shows that every unification scheme as
superstrings, supergravity or grand unified theories, gives
effective actions where non-minimal couplings to the geometry or
higher order terms in the curvature invariants come out (an
example is the above axion coupling). Such contributions are due
to one-loop or higher-loop corrections in the high curvature
regimes near the full (not yet available) quantum gravity regime
\cite{odintsov}.  However, in the weak-limit approximation, all
these theories should be expected to reproduce the Einstein
General Relativity which, in any case, is experimentally tested
only in this limit \cite{will}.

This fact is matter of debate since most of relativistic theories
{\it do not} reproduce Einstein results at the Newtonian
approximation but, in some sense, generalize them \cite{grgrev}
giving Yukawa-like corrections to the Newtonian potential which
could have interesting physical consequences.

For example, some authors  have shown that a conformal theory of
gravity is nothing else but a fourth-order theory containing such
terms in the Newtonian limit and, by invoking these results, it
could be possible to explain the missing matter problem ''without"
dark matter \cite{mannheim}.

A  general effective theory of
gravity coming from the renormalization of quantum fields on curved space-time can have an effective action of the form
 \begin{eqnarray}
  \label{extended} {\cal A}=\int &&
d^{4}x\sqrt{-g}\left[F(R,\Box R,\Box^{2}R,..\Box^kR,\phi)+\right.\nonumber\\
& & \left. -\frac{\epsilon}{2}
g^{\mu\nu} \phi_{;\mu} \phi_{;\nu}+ {\cal L}_{m}\right], \end{eqnarray}
 where $F$ is
an unspecified function of curvature invariants $R$ and $\Box R$ and of  scalar
fields $\phi$. The term ${\cal L}_{m}$ is the minimally
coupled ordinary matter contribution. Actually,
 more complicated invariants like $R_{\mu\nu}R^{\mu\nu}$,
$R_{\mu\nu\alpha\beta}R^{\mu\nu\alpha\beta}$,
$C_{\mu\nu\alpha\beta}C^{\mu\nu\alpha\beta}$  are also
possible but we do not need them for the present purposes.

The weak field limit of theories of the form lead to gravitational potentials of the form  \cite{mnras,stelle,correction,odibox}
\beq \label{yukawa}
V(r)=-\frac{G_{\infty}m}{r}\left[1+\sum_{k=1}^{n}\alpha_k
e^{-r/\lambda_k}\right]\,,
\eeq
where $G_{\infty}$ is the value of
the gravitational constant as measured at infinity, $\lambda_k$ is
the interaction length of the $k$-th component of non-Newtonian
corrections. The amplitude $\alpha_k$ of each component is
normalized to the standard Newtonian term (see
\cite{will,principia} for further details). The discussion
involves also the variation of the gravitational coupling. As an
example, let us take into account only the first term of the
series in $(\ref{yukawa})$ which is usually considered the leading
term (this choice is not sufficient if other corrections are
needed). We have

 \beq
\label{yukawa1} V(r)=-\frac{G_{\infty}m}{r}\left[1+\alpha_1
e^{-r/\lambda_1}\right]\,. \eeq The effect of non-Newtonian term
can be parameterized by $(\alpha_1\lambda_1)$. For large
distances, at which $r\gg\lambda_1$, the exponential term
vanishes and the gravitational coupling is $G_{\infty}$. If
$r\ll\lambda_1$, the exponential becomes unity and, by
differentiating, we get

\beq \label{yukawa2}
G_{lab}=G_{\infty}\left[1+\alpha_1\left(1+\frac{r}{\lambda_1}\right)e^{-r/\lambda_1}\right]\,, \eeq
where
$G_{lab}=6.67\times10^{-8}$ g$^{-1}$cm$^3$s$^{-2}$ is the usual
Newton constant measured by Cavendish-like experiments. Of
course,  $G_{\infty}$ and $G_{lab}$ coincide in standard gravity.
It is worthwhile to note that, asymptotically, the inverse square
law holds but the measured coupling constant differs by a factor
$(1+\alpha_1)$. In general, any exponential correction introduces
a characteristic length that acts at a certain scale for the
self-gravitating systems. For a discussion and applications to the large scale structure, see \cite{mnras}.

This approach has been pursued by some authors who  tested
non-Newtonian corrections by ground-based experiments using
totally different approachess \cite{fischbach,speake,eckhardt1}.
The general outcome of these experiments, even retaining only the
term $k=1$, is that a ''geophysical window" between the laboratory
and the astronomical results has to be taken into account. In
fact, the range

\beq |\alpha_1|\sim 10^{-2}\,,\qquad \lambda_1\sim 10^2\div
10^3\,\mbox{m}\,,\eeq is not excluded at all. The sign of
$\alpha_1$ tells us if corrections are attractive or repulsive.
Another interesting suggestion has been given by Fujii
\cite{fujii1}, which proposed that the exponential deviation from
the Newtonian standard potential (the ''fifth force") could arise
from the microscopic interaction which couples to nuclear isospin
and baryon number.

The astrophysical counterpart of these non-Newtonian corrections
seemed ruled out till some years ago due to the fact that
experimental tests of general relativity predict ''exactly" the
Newtonian potential in the weak energy limit, ''inside" the Solar
System. Recently, as we said above,
 indications of an anomalous, long--range acceleration
revealed from the data analysis of Pioneer 10/11, Galileo, and
Ulysses spacecrafts makes these Yukawa--like corrections come
into play \cite{anderson}. Besides,
reproduced the flat rotation curves of spiral galaxies can be reproduced by assuming \cite{sanders}
\beq \alpha_1=-0.92\,,\qquad \lambda_1\sim 40\,\mbox{kpc}\,.\eeq
The main hypothesis is that the additional gravitational
interaction is carried by an ultra-soft vector boson whose range
of mass is $m_0\sim 10^{-27}\div 10^{-28}$eV. The action of this
boson becomes efficient at galactic scales without the request of
enormous amounts of dark matter to stabilize the systems.

On the other hand, by asking for a characteristic length emerging
from the standard theory of cosmological perturbation, it is
possible to explain the observed segregation of hot stellar
systems in the so called {\it fundamental plane} of galaxies
(''ordinary" and ''bright" galaxies) \cite{capaccioli}. In that
case, the length is the ''Jeans length" of the protogalaxy
($\lambda\sim 3\div 10$ kpc) and, due to this characteristic
size,  a Yukawa correction was found in the gravitational
potential with a characteristic interaction ''length" of the same
order of magnitude of that proposed by Sanders.

The emergence of characteristic
lengths could be addressed by the above very fundamental approach capable of giving
rise, in principle, to all  self-gravitating systems. Our guess is
that such lengths give rise to non-Newtonian corrections in the
gravitational potential and then the aggregation of systems results
extremely natural without the "ad hoc" introduction of dark matter
constituents.

In the next section, we will discuss what we intend for the
characteristic size of a self-gravitating system and then how they
can be achieved starting from primordial quantum black holes in
the framework of conformal theories.

\section{Astrophysical self-gravitating systems}
\label{sei}

 In general, the concept of ''size" of a self-gravitating system
is not well-based since, in several cases, the boundary cannot be
univocally defined. Let us briefly define globular clusters,
galaxies, clusters and super-clusters of galaxies considering
 their typical lengths and masses. This point is crucial since we
are taking into account only systems where gravity is the only
overall interaction acting between the components. In this sense,
a star is not a purely self-gravitating system since, inside it,
gravity is balanced by the pressure due to electromagnetic and
nuclear interactions. However, we can take into account stars as
granular constituents of globular clusters and galaxies and define
a typical interaction length as the ''size"  of a planetary system
around a star.

A  globular cluster is a very compact self-gravitating stellar
system whose typical radius is $R_{gc}\sim 10$ pc. It contains up
to $10^6$ stars ($M_{gc}\sim 10^6 M_{\odot}$) and is assumed
completely virialized due to collisional interactions between
stars.

On the other hand, a galaxy is a collisionless, diffuse
gravitating system without an effective boundary. Astronomers
define operative characteristic sizes as the {\it effective
radius} $R_e$ which is the radius of the isophote containing half
of the total luminosity, or the {\it tidal radius} $R_t$
corresponding to the distance from the center where the density
drops to zero \cite{binney,vorontsov}. Other definitions are
possible by using photometry or kinematics but, assuming as a
typical interaction size a length $R_g\sim 1\div 10$ kpc is quite
reasonable from dwarf to giant galaxies. It is important to point
out that several authors assume that the halo of giant galaxies
can extend as far as $100$ kpc from the center taking into account
the dark matter component. Here we do not assume any dark matter
hypothesis and do not want to enter into details of galactic
dynamics and morphology. For example, Milky Way, a typical spiral
galaxy, has an observed  disk scale length $R_d\simeq 3.5\pm 0.5$
kpc while kinematics of globular clusters and $21$ cm-radio
observations of neutral hydrogen give a maximal halo extension of
$20\div 30$ kpc. For our purposes, assuming $10$ kpc as a
characteristic size, with a possible error of an order of
magnitude, is a good number. Typical masses are $M_g\sim
10^{10\div 12}M_{\odot}$ for giant galaxies and $M_g\sim 10^{8\div
9}M_{\odot}$ for dwarf galaxies.

As a  cluster of galaxies, following Abell \cite{abell}, we define
a self-gravitating system whose granular components are galaxies
with a typical radius $R_{cg}=R_a\simeq 1.5 h^{-1}$ Mpc (the Abell
radius) and a typical mass $M_{cg}\sim 10^{15} h^{-1}M_{\odot}$
for rich clusters, where $h$ is the dimensionless Hubble constant
whose value is in the range $0.5<h<1$ \cite{peebles}.

Finally, a super-cluster is a self-gravitating system of clusters
of galaxies whose typical size is $R_{sc}\sim 10\div 100 h^{-1}$
Mpc and typical mass is $M_{sc}\sim 10^{15\div 17}
h^{-1}M_{\odot}$.

 Groups of galaxies are systems containing
$10\div 20$ galaxies, as our Local Group, but there are no
evidences that they could be considered self-gravitating systems
and, in any case, they are always part of  more extended cluster
of galaxies (in the case of Local Group, it is a part of the
Virgo Cluster).

The main difference between a globular cluster and the other
systems is that the former is a collisional system while the
others are collisionless. This fact implies a completely different
dynamical treatment \cite{binney}.

The properties of these self-gravitating systems can be deduced by
assuming them to be relaxed and virialized systems where gravity
is the only overall interaction \cite{binney}. This assumption
 is, some times, not completely justified. In
fact, we have to keep in mind that these systems undergo
environmental effects, being never completely isolated; they
always belong to larger gravitationally bound systems and the
observational times are so short that the overall dynamics can
only be extrapolated \cite{binney,vorontsov}.

Furthermore, as we said, the dynamics of astrophysical systems
have to be related to  cosmological evolution  so that, in today
observed dynamics, some quantum signature  of  primordial quantum
perturbations should be present \cite{sakharov}.

However the main difficulty, is to provide a physical route
connecting the sizes of astrophysical structures with the
extremely small numbers of quantum mechanics.

As a first step, we can  build a model of the above structures
composed of self-gravitating microscopic constituents (nucleons)
 which undergo quantum fluctuations coming from a very fundamental mechanism.

We introduce as the only observational input the number of
nucleons contained.  The characteristic dimension of such a model,
as supposed above, is a functions of the microscopic nucleon
scales (the Compton wavelength of a nucleon  $\lambda_c\sim
10^{-13}$ cm) and of the number of microscopic constituents.

The result is that  the characteristic radii so deduced,
numerically, coincide with those observed considering the usual
gravitational constituents  (stars for globular clusters and
galaxies or galaxies themselves for clusters and super-clusters
of galaxies, not nucleons).

Besides, we obtain a  scaling relation between the  units of
length and action of the  granular gravitational components,
ranging from nucleons up to stars and galaxies.

On the other hand, the characteristic dimensions of astrophysical
structures appear to be independent of the scale of the
constituents considered. It is only needed that they depend on a
minimal scale of length, which is, in order of magnitude,  the
Compton wavelength of a nucleon naturally deduced from the Planck
length (see the above discussion).

These results  suggest a macroscopic quantum coherence for large
scale gravitational systems.

Furthermore, the emergence of these characteristic scales could
have a dynamical counterpart in the  Yukawa corrections of
gravitational potential as in Eq.(\ref{yukawa}).

\section{Self-gravitating systems from a characteristic fundamental length}
\label{sette}
We want to show now that any self-gravitating system (from a star
up to the whole Universe) can be achieved by quantum black holes
defining a characteristic fundamental length

We have to consider  the total action for the bound system with a
very large number $N$ of constituents. The working hypothesis is
that such an action can be achieved from the Planck constant.  Let
$E$ be the total energy. Let ${\cal T}$ be the characteristic
global time of the system (e.g. the time in which a particle
crosses the system, or the time in which the system evolves and
becomes relaxed or virialized). By these two quantities, we get

 \beq
\label{1} A \cong E{\cal T} \, , \eeq which is the total action of
the system (see also Eq.(\ref{J1}) to see how the heuristic following discussion  can be related to  fundamental arguments) .
The system could undergo a time-statistical fluctuation, so that the
characteristic time $\tau$ for the stochastic  motion per particle
is
\beq \label{2} {\tau}\cong\frac{{\cal T}}{\sqrt{N}} \, . \eeq
We can then define an energy per granular component

\beq \label{3} \epsilon\cong\frac{E}{N} \, , \eeq so that the
characteristic (minimal) unit of action $\alpha = \epsilon \tau$
per granular component is expressed by the  scaling relation

\beq \label{4} \alpha=\epsilon\tau\cong\frac{A}{N^{3/2}} \, . \eeq
This general considerations become physically consistent if the
above fundamental theory is taken into account.  Let us now
consider the observational data for the above self-gravitating
systems, in order of magnitude. Since we are taking into account
virialized systems, we can assume

\beq \label{vir} 2E_{k}+U=0\,, \eeq
 where $E_{k}$ is the kinetic energy and $U$ the gravitational
energy. The total energy $E$ then is

\beq\label{total} E\simeq E_{k}\simeq N M v^2 \eeq where $N$ is
the typical number of granular components (e.g. stars in a galaxy
or in a globular cluster, or galaxies in a cluster or
super-cluster of galaxies;  $M$ is the typical mass (e.g.
$1M_{\odot}$ for a Main Sequence star or $10^{10\div 11}M_{\odot}$
for a galaxy like Milky Way);
 $v$ is a characteristic typical velocity
which we choose to be the circular speed of the stars in the disk
of galaxies $(\simeq 10^{7\div 8}$cm/sec)  or the velocity
dispersion of the galaxies inside a cluster $(\simeq 10^{8\div
9}$cm/sec). All these numbers are quite accurately measured by the
methods of stellar kinematics, stellar statistics and photometry
\cite{binney}.
 All the above quantities entering into the definitions of the
energy, time and action scales
 are   quantities coming from observations. In particular, nowhere we
introduce the characteristic radius $R$ of the structures since
this is what we wish to predict in the framework of our approach.

Let us take into account a galaxy. The energy per unit of mass is
of the order $10^{15}$ (cm/sec$)^2$, while the period of a
galactic rotation, which can be assumed as the characteristic
global time, is of the order \cite{binney}

\beq \label{5} {\cal{T}}_{rot} \cong 10^{15}\mbox{sec} \, , \eeq
and finally the total mass of a typical galaxy is of the order
\cite{binney}

\beq \label{6} M_g=N_{s}^{(g)}M_s \cong 10^{44}\mbox{gr} \, . \eeq
From Eq.(\ref{1}), combining these numbers, we get the typical
action

\beq \label{7} A \cong 10^{74}\mbox{erg sec} \, . \eeq The typical
number of nucleons in a galaxy is \cite{binney}

\beq \label{8} N_{p}^{(g)} \cong 10^{68} \, . \eeq
 Inserting these numbers
in Eq.(\ref{4}) we see that, up to an order of magnitude, the
characteristic unit of action $\alpha$ of a  galaxy, considered
as an aggregate of nucleons, is of the order of the Planck action
constant, $h \sim 10^{-27}$ erg sec. It is interesting to notice that such a result is in full agreement with the quantization relation (\ref{VII}).
In fact using in (\ref{VII})
\begin{equation}
M_{g}=N_{p}^{(g)}m_{p}=10^{68}\times10^{-24}g=10^{44}g\,,
\end{equation}
we  get
 \begin{equation}
 n_g=\frac{G [N_{p}^{(g)}m_p]^2}{\hbar c}=10^{98}\,,
 \end{equation}
  to be compared with
  \begin{equation}
  (N_{p}^{(g)})^{3/2}=10^{102}
 \end{equation}
   see Eqs. (\ref{4}) and (\ref{8}) . That is the  two results fully agree, within a few orders of magnitude allowed from statistical fluctuation, due for example for the presence of  a "small" amount ($\simeq 2\%$) of heavy elements in the Universe (that is elements heavier than cosmological helium which could give rise to fluctuations in the nucleosynthesis \cite{burles}). It is worth  stressing
that also if dark matter is considered, the result does not change
dramatically since the mass to luminosity ratio is of the order
$10 \div 100$.

As a further step, we note that Eq.(\ref{4}), together with the
numerical result $\alpha \cong h$, can be re-formulated as a
scaling relation for the {\it mean} action per microscopic
component $a \equiv A/N$, that is

\beq \label{9} a \cong h \sqrt{N} \, . \eeq We can then deduce
that the fluctuative factor $\sqrt{N}$ provides the rescaling
coefficient from the microscopic fundamental scales to the
characteristic macroscopic dimensions.
Let us now take into account the lengths. Given the nucleons as
the basic microscopic constituents in our model, the natural
quantum unit of length associated to each single constituent is
the Compton wavelength $\lambda_c = h/mc$, with $c$ the velocity
of light, and $m \cong m_{p} \cong 10^{-24}$ gr, the proton mass.
 In analogy with
Eq.(\ref{9}), we have, in general,

\beq \label{10} R \cong \lambda_{c} \sqrt{N} \, . \eeq For a
galaxy, with $N$ given by Eq.(\ref{8}),  we obtain

\beq \label{11} R_{g} \cong 10^{21} \div 10^{22} \mbox{cm} \simeq
1 \div 10 \mbox{kpc} \,  \eeq which as we said above is a length
of a galaxy. In particular, the numerical agreement of
Eq.(\ref{10}) with the observed galactic radii, is interesting,
independently of the present derivation, since it links the scale
of a large structure like a galaxy to the Compton wavelength of
the elementary constituents (the nucleons) and to the total
number of such constituents. It is worth noticing that, for typical galaxies, $R_{g}$ is the
characteristic dimension where their rotation curve can be assumed
flat \cite{binney} and where the halo and the disk stabilize each
other.

The validity of Eq.(\ref{10}) is not restricted to the galaxies,
but provides the correct order of magnitude of the observed radii
also if one considers the other structures which we mentioned
above.

In the case of globular clusters, considering the suitable $N$
(which we easily deduce by the number of stars which constitute
them, i.e. $10^6$), we get

\beq \label{11a} R_{gc} \cong 10^{18} \div 10^{19} \mbox{cm}
\simeq 1 \div 10 \mbox{pc} \, . \eeq

For clusters of galaxies we obtain $R_{cg}\cong 1$ Mpc and for
super-clusters $R_{sg}\cong 10\div 100$ Mpc.

The discussion can be extended to the whole Universe, and to other
astrophysical objects, such as planetary systems  (like the Solar
System), provided one inserts in Eq.(\ref{10}) the correct value
of the number of nucleons $N$ contained in such structures.

These findings indicate that the quantum parameter $\lambda_{c}$
and  the number of nucleonic constituents $N$, determine the
observed astrophysical and cosmological dimensions.

 The crucial objection to these results would be
that stars or single galaxies, rather than nucleons, are the
natural candidates as elementary gravitational constituents of a
typical galaxy or a typical cluster of galaxies.

This apparent difficulty can be solved deriving a simple scaling
law, which holds true at any scale. Let us take into account, for
example, the number of stars $N_{s}^{(g)}$ contained in a typical
galaxy, and the number $N_{n}^{(s)}$ of nucleons in a star. We
can then obviously write, for the total number of nucleons in a
typical galaxy

\beq \label{15} N \cong N_{s}^{(g)}N_{n}^{(s)} \, . \eeq By
Eq.(\ref{15}), we can write Eq.(\ref{10}) as

\beq \label{16} R_g \cong \lambda_{s} \sqrt{N_{s}^{(g)}} \, , \eeq
where

\beq \label{17} \lambda_{s}  \equiv \frac{A_{s}}{M_{s}c}\,,\quad
A_{s} \equiv  h [N_{n}{^{(s)}}]^{3/2} \,,\quad M_{s} \equiv m
N_{n}^{(s)} \, . \eeq Here, as above, $M_{s}$ is  the total mass
of a star, while the quantity $A_{s}$ is the characteristic unit
of action of a star in the framework of our model, taking the
stars as the elementary constituents of a typical galaxy.
Inserting the numerical values \cite{binney} $N_{n}^{(s)} \cong
10^{57}$, $N_{s}^{(g)} \cong 10^{10} \div 10^{12}$, we obtain

\beq \label{18} \lambda_{s} \cong 10^{13} \div 10^{15} \mbox{cm}
\, , \eeq
 which agrees with the typical range of interaction of a star
(e.g. that of the Solar System), while for $R_g$ we obviously
obtain again the value (\ref{11}).

Therefore, Eqs.(\ref{10}) and (\ref{16}) show that we can derive
the observed galactic radius $R_g$ either by considering a galaxy
as a gas of $N$ nucleons with the fluctuation (\ref{2}) defined
with respect to $N$, or by considering, as usual, a typical
galaxy as a gas of stars and assuming the fluctuative ansatz
(\ref{2}) rescaled with respect to the number of stars
$N_{s}^{(g)}$.

The reason for the validity of this relation  (which, in
principle, holds on any scale)  relies on the existence of a
minimal scale of action which is needed for mechanical stability.
In fact, the numerical value of the unit of action $A_{s}$
defined in Eq.(\ref{17}) is $\cong 10^{58}$ erg s and thus
coincides, in order of magnitude, with the total action for a
typical star. Thus the rescaling relations (\ref{4}) and
(\ref{10}) hold true also for a star, and $\lambda_{s}$ appears
as the effective macroscopic ``Compton wavelength'' of a star.
However $\lambda_s$ is the typical range of interaction also in
the case of a globular cluster giving
$R_{gc}\cong\lambda_{s}\sqrt{N_{s}^{(gc)}}\sim 1\div 10$ pc.

Immediately we derive $\lambda_g$ as the range of interaction for
galaxies considered as granular constituents of clusters and
super-clusters. Analogously, we have

\beq \label{17a} \lambda_{g}  \equiv \frac{A_{g}}{M_{g}c}\,,\quad
A_{g} \equiv  h [N_{n}{^{(g)}}]^{3/2} \,,\quad M_{g} \equiv m
N_{n}^{(g)} \, . \eeq and then $\lambda_g\cong 10\div 100$ kpc.
In this case, we can hierarchically consider a cluster or a
super-cluster of galaxies as a gas of nucleons, stars or galaxies.
It is interesting to note that $\lambda_g$ is the observed typical
separation length between galaxies in a cluster.

At this point, it is straightforward the  connection to the
non-Newtonian gravitational potential (\ref{yukawa}). As we
discussed above, the Yukawa corrections have to be reconducted to
the emergence of typical scales for self-gravitating systems. For
example, as discussed in \cite{sanders} and in \cite{eckhardt},
  by the interaction ranges of some vector bosons, it is possible
  to explain the flat rotation curves of spiral galaxies without
  asking for large amounts of dark matter. Yukawa corrections naturally
  emerge in relation to these interaction ranges. The main shortcoming of
  their approach is that, till now, no ultra-light vector boson has
  been detected and the requested interaction lengths $\lambda\sim
  10$ kpc are very hard to justify.

  By our  hypothesis, as we discussed above,
  $\lambda_g\sim 10$ kpc naturally emerges by taking into
  account  the granular components of a
  galaxy. Using, Eq.(\ref{yukawa1}) where we assume
  $\lambda_1=\lambda_g$, the arguments in \cite{sanders} and in
  \cite{eckhardt} are easily recovered.

  Besides, the anomalous, long-range acceleration reported in
  \cite{anderson} immediately outside the Solar System, could be explained
  considering a Yukawa correction in the Newtonian potential related to
  a length as  $\lambda_s$ which
  can be considered as  the typical range of interaction of a star as the
  Sun (a system with gravitationally bound planets).

\section{\normalsize \bf Discussion and Conclusions}
\label{otto}

In this paper, we have discussed the possibility that the
characteristic sizes of astrophysical self-gravitating systems
could be deduced by scaling laws relating the observed
macroscopic dimensions to the microscopic fundamental scales originated by the primordial quantum black holes.
The net effects at macroscopic scales are Yukawa-like
corrections to the Newtonian potential which become relevant  in
the range $r\sim \lambda$. This fact could be connected to the
well-known absence of screening mechanisms for gravity
\cite{bertolami}.

Before drawing the conclusions, we have to discuss the scales of
action involved. Another link between the quantum unit of action
$h$ and the cosmological scales is provided by the so--called
Eddington--Weinberg relation $h \cong G^{1/2}m^{3/2}R^{1/2}$,
where $G$ is the Newton gravitational constant, $m$ is the mass
of the nucleon, and $R$ is the radius of the Universe.
 If one takes for $R$ the various definitions of cosmological radius (Hubble
radius, causal radius, or last scattering radius) \cite{peebles}
which range from $R=10^{26} cm$ to $R=10^{30} cm$, one obtains a
value for the unit of action $\alpha$ ranging from $10^{-26}$ erg
sec to $10^{-27}$ erg sec which is usually  assumed to coincide
with the Planck constant $h$ in order of magnitude \cite{barrow,roy}.

A similar relation can be deduced also for the self-gravitating
structures which we have discussed, if one inserts in the
equations for $A$ and $\alpha$ the gravitational energy $U(R)$
and a characteristic gravitational time ${\cal T}$ needed for the
relaxation of the system.
 In this case, a sort of
Eddington--Weinberg relation can be derived and appears to hold
also for galaxies, and other large scale structures as clusters
and super-clusters of galaxies yielding a microscopic unit of
action of the order of $10^{-(27\div 28)}$ erg sec.

The numerical coincidence up to two orders of magnitude of the
Eddington--Weinberg relations for the Universe and for large scale
structures is not purely accidental if the quantum black holes are
ruling the astrophysical structures. Therefore, as we have shown,
what is really significant in our model, both on astrophysical and
cosmological scales, are the micro/macro scaling relations and
connectivity factors, that is Eqs.(\ref{2}), (\ref{4}), and
(\ref{10}), which hold true for all systems and are built starting
from the basic assumption of nuclear granularity \cite{scott1,scott2}.

In conclusion, the typical hierarchical  sizes of astrophysical
structures could be explained by taking into account a very
fundamental length giving rise to the  typical interaction ranges
of self-gravitating systems. Dynamics is implemented by a
non-Newtonian gravitational potential where Yukawa corrections
effectively act at that given typical scale. However,  the value
of the gravitational coupling is different at the various
distances depending on the interaction ranges $\lambda_c,
\lambda_s,\lambda_g$. Finally, we want to stress that  no additional (unknown) dark matter particle has been introduced and
the standard nucleons can completely account for sizes and
stability of astrophysical structures.

\end{document}